\documentclass[aip,reprint]{revtex4-1}

\usepackage[T1]{fontenc}
\usepackage{lmodern}
\usepackage{graphicx}
\usepackage{subfigure}
\usepackage{dcolumn}
\usepackage{bm}
\usepackage{amsmath}
\usepackage{color}
\usepackage{ulem}

\def\Cint{C$_{\rm int}$}

\def\deg{$^\circ$}

\begin{document}

\title{Tetrahedral Bonding in Twisted Bilayer Graphene by Carbon Intercalation}
\author{Anindya Roy}
 \email{anindyar@mit.edu}
\affiliation{
Materials Science \& Engineering, Johns Hopkins University, Baltimore, MD 21218, USA}
\altaffiliation{
Present address: Office of Digital Learning, Massachusetts Institute of Technology, Cambridge, MA 02139, USA}

\author{Hridis K.~Pal}
 \email{hridis.pal@physics.gatech.edu}
\affiliation{Laboratoire de Physique des Solides, CNRS UMR 8502, Universit\'{e} Paris-Sud, Universit\'{e} Paris-Saclay, 91405  Orsay Cedex, France}

\date{\today}

\begin{abstract}
Based on {\it ab initio} calculations, we study the effect of intercalating twisted bilayer graphene with carbon.
Surprisingly, we find that the intercalant pulls the atoms in the two layers closer together locally when placed in certain regions in between the layers, and the process is energetically favorable as well. This arises because in these regions of the supercell, the local environment
allows the intercalant to form tetrahedral bonding with nearest atoms in the layers. Intercalating AB- or AA-bilayer graphene with carbon does not produce this effect; therefore, the nontrivial effect owes its origin to
both using carbon as an intercalant and using twisted bilayer graphene as the host. This opens new routes
to manipulating bilayer and multilayer van der Waals heterostructures and tuning their properties in an
unconventional way.
\end{abstract}


\maketitle

\def\scr{\scriptsize}

Following the successful separation of graphene as the first truly 2D crystal \cite{geim1}, the list of experimentally realized 2D materials has grown steadily \cite{geim2}. With the proliferation of such new 2D crystals, a new field has emerged---that of 2D van der Waals heterostructures \cite{van1,van2,van3,pal,palarxiv}. By stacking 2D layers on top of each other, one can create heterostructures with new properties and functionalities that are absent in both the parent layers as well as available 3D materials. A common feature among these systems is that the layers are held together by weak van der Waals force; therefore, there is a strong asymmetry in the in- and out-of-plane coupling between atoms. This often leads to the layers being misaligned, giving rise to large moir\'e superlattices. The simplest example is twisted bilayer graphene (TBG)---two layers of graphene mutually rotated by an an arbitrary angle---other examples being graphene on hBN, bilayer dichalcogenides, etc. 

\begin{figure}[ht!]
\includegraphics[width=3.0in]{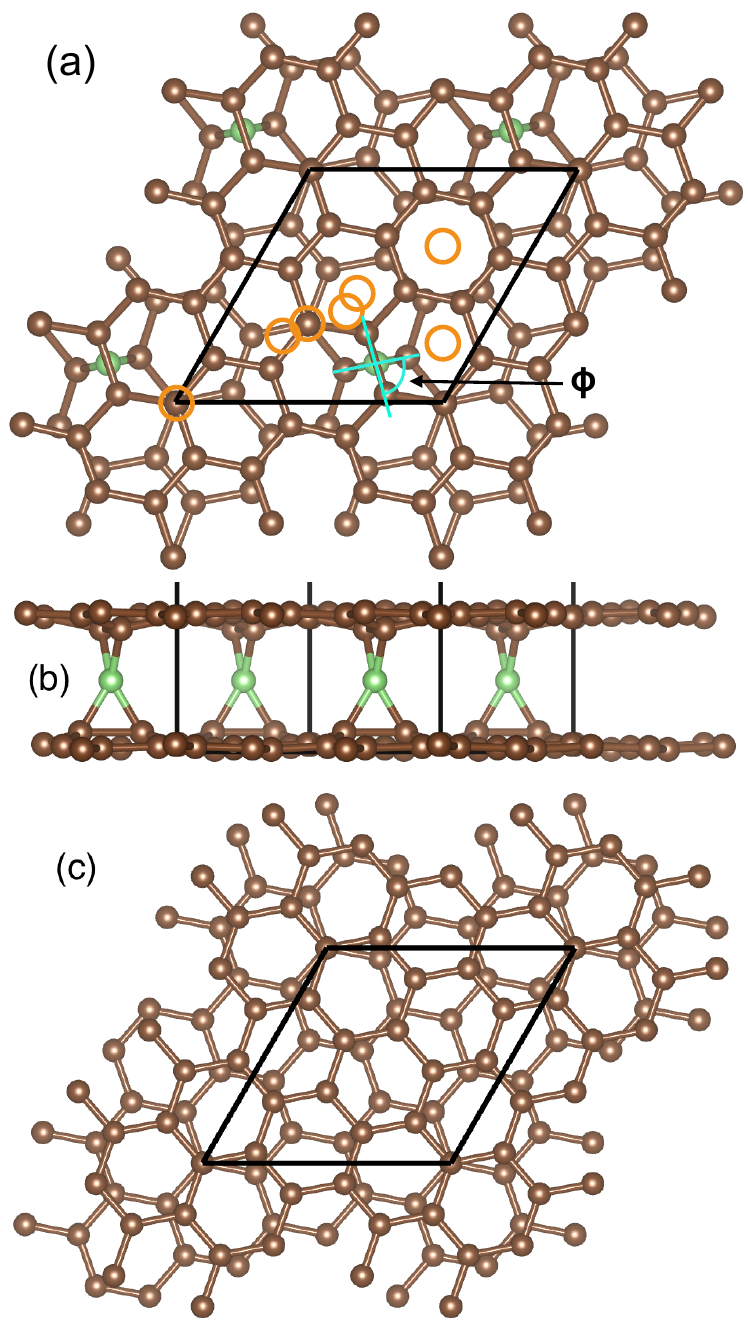}
\caption{\label{fig:c_int} (Color online) Top view (a) and the side view (b) of TBG (38.21\deg) with the intercalated C atom in green. (c) Top view of TBG (21.79\deg). The black rhombuses outline the unit cells, and only the second-from-right intercalant is in the unit cell in panel (b). In panel (a), orange circles refer to the additional initial in-plane positions (halfway between the layers) we investigated for the C intercalants. In the same panel, we also show the angle $\phi$, defined in the text, via two intersecting straight-line segments situated on top of the green intercalant. It can be seen that, while in (a) local arrangements of atoms in the layers allow for tetrahedral arrangement, no such arrangement is seen in (c).}
\end{figure}

\begin{figure}
\includegraphics[width=3.0in]{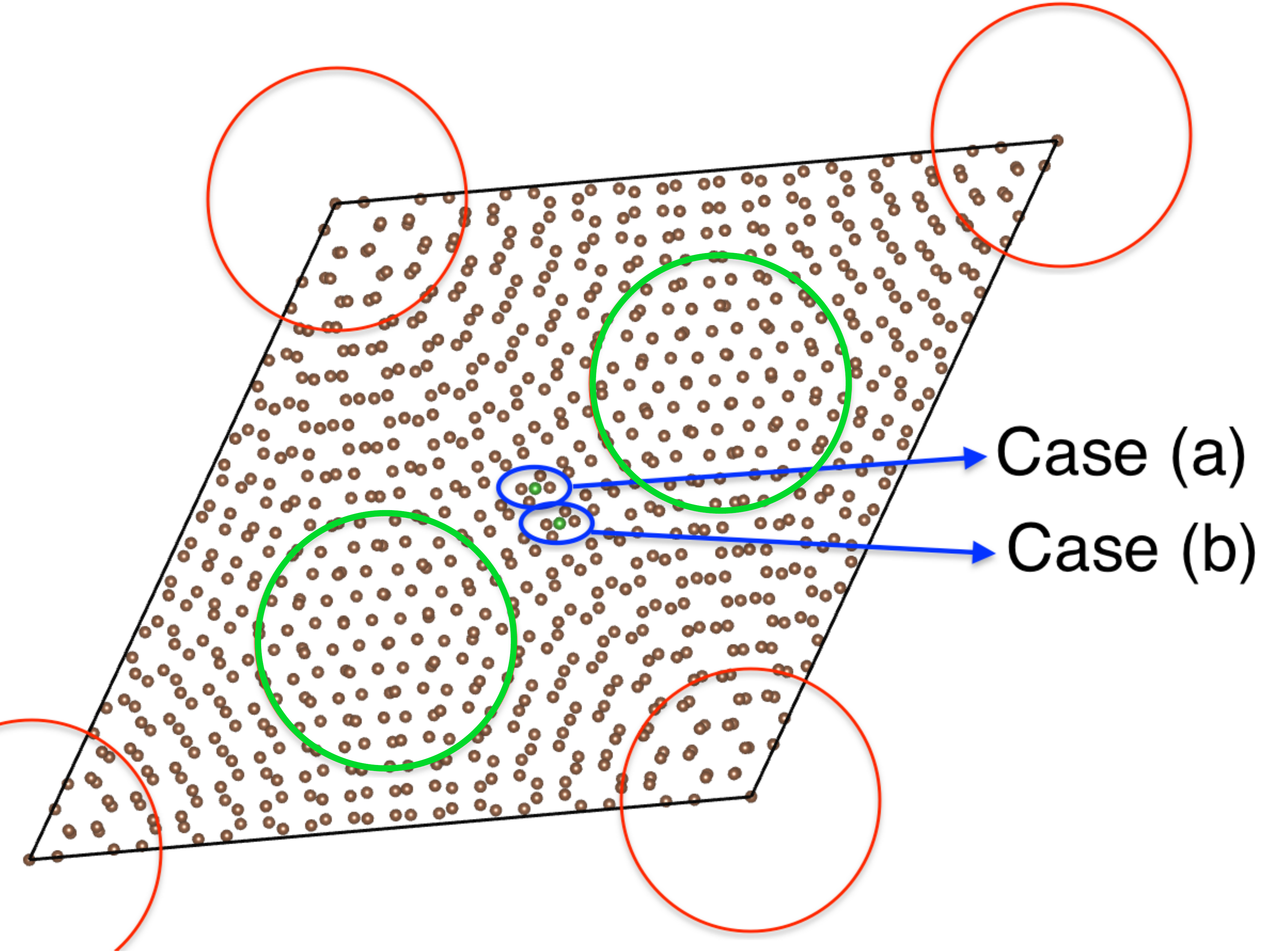}
\caption{\label{fig:m3-89} (Color online) TBG (3.89\deg) unit cell with two possible positions for intercalation [cases (a) and (b)] referred to in Table~\ref{table:compcell} where it is locally tetrahedral. The green circles (inside the unit cell) show locally AB- and BA-like regions, and the red circles (at the vertices of the unit cell) show locally AA-like regions.}
\end{figure}

At the heart of such new behavior in these heterostructures is the presence of new energy and length scales---the interlayer coupling and moir\'e lattice constant, respectively \cite{lopes,bistritzer,shallcross,hofstadter1,hofstadter2}.  
In addition to providing a new length scale, 
the moir\'e superlattice also introduces new local features in the system. The atomic registry of the two layers changes continuously  within a supercell, providing new local environments in a system not found in parent 2D or other 3D materials. For example, in AB-bilayer graphene, there are only two kinds of local environments: an atom with another atom on top, and an atom with a void on top; in TBG, however, many more local environments are possible, increasing in number as the size of the superlattice increases (see Figs.~\ref{fig:c_int} and \ref{fig:m3-89}). Local perturbations are therefore expected to affect such systems differently compared to the usual AB- or AA-bilayer. 

One way of introducing such local perturbations is through intercalation. Owing to the weak van der Waals forces between the layers, it is natural to expect that the intercalant will push away the atoms of the host layers, thereby increasing the overall interlayer separation. Indeed, this is borne by existing studies in traditional layered materials, e.g., graphite \cite{palintercal} and in TBG  as well \cite{TBGintercal1}. In this letter, based on {\it ab initio} calculations, we show that, surprisingly, carbon intercalation in TBG is capable of producing the opposite effect: the intercalant atom pulls the nearby atoms in the two layers closer together by a substantial amount! On closer inspection of such cases, we find that this type of intercalation arises in regions of the supercell where the local environment allows the intercalant to form tetrahedral bonding with nearest atoms in the layers. Although the effect is usually confined locally around the intercalant, we observe global reduction in average interlayer separation in case of the smallest  TBG unit cell (corresponding to the twist angle 38.21\deg). Moreover, cases where we observe tetrahedral bonding with the intercalants also show the largest energy gain from intercalation, when compared to other intercalation positions studied. To the best of our knowledge, this is the first time such a behavior has been observed. Intercalating AB- or AA-bilayer graphene with carbon does not produce this effect. Hence, this unusual behavior owes its origin to both using carbon as an intercalant and using TBG as the host.  Such an unconventional behavior opens new routes to manipulating bilayer and multilayer van der Waals heterostructures and tuning their properties in a novel way.

We study the above effects in TBG using density functional theory and related methods. The DFT calculations presented in this work are performed with {\scr Vienna Abinitio Simulation Package} ({\scr VASP})\cite{kresse-vasp1, kresse-vasp2}. The van der Waals (vdW) forces are important to describe 2D layered structures such as bilayer graphene, and there are a number of available schemes to account for vdW forces in the DFT framework. One such scheme---DFT-D2 or Grimme's method\cite{Grimme2006}---achieves this via a parameterized method. The DFT-D2 method has been used previously with the Perdew-Burke-Ernzerhof (PBE) parametrization of the generalized gradient approximation exchange-correlation functionals \cite{perdew1996phys, perdew1997emission} to describe 2D-layered materials, including bilayer graphene\cite{caciuc_ab_2012, Lee2012, van_der_zande_tailoring_2014}. On the other hand, local density approximation (LDA) exchange-correlation functionals spuriously produce close-enough structure parameters for these cases without explicitly accounting for vdW forces, as detailed in Ref.~\onlinecite{rydberg_hard_2003} and references therein, and as is often used to study bilayer graphene and TBG\cite{gong_first-principles_2010, landgraf_electronic_2013, ohta_evidence_2012}. For our calculations, we verified on test cases that the PBE+DFT-D2 scheme produces similar values for the calculated lattice parameters, interlayer separation, and energy gain arising from intercalation, and decided to use the LDA-based projector-augmented wave (PAW) pseudopotentials\cite{Blochl-paw1994}.

The self-consistent calculations (SCF) and ionic relaxations with fixed lattice parameters have a plane-wave cutoff of 400 eV, whereas a higher cutoff of 520 eV is applied when optimizing cell parameters. An $11\times11\times1$ Monkhorst-Pack $k$-point mesh is used for the Brillouin zone integration for the 4-atom unit cell of AA- or AB-stacked bilayer graphene, and proportionally smaller $k$-point meshes are used for the larger TBG unit cells. Forces on ions are converged to less than 0.005~eV/\AA. A vacuum layer of $\approx12.5~\AA$ is used to separate the bilayer structure from its periodic images in the out-of-plane direction.

\begin{table}
\caption{\label{table:compcell} Moir\'e cell results: Columns 1 and 2 record the cell parameters and the twist angle $\theta$ of the TBG, columns 3 and 4 the energy gain on intercalation and the average interlayer separation, column 5 the average C-C$_{\rm int}$ bond length, and columns 6 and 7 the angles before and after ionic relaxation ($\phi_{\rm init}$ and $\phi_{\rm fin}$), respectively.}
\begin{ruledtabular}
\begin{tabular}{llccccc}
Cell & Twist & $\Delta$E & Layer & C-C$_{\rm int}$ & \multicolumn{2}{c}{Angle $\phi$}\\
param. & angle &  &  sep. & avg. BL & init. & fin. \\   
(\AA) & (Degree) & (eV)  & (\AA) & (\AA) & (Degree) & (Degree) \\
\hline
\phantom{1}6.502 & 38.21  & 5.283 &  3.157 & 1.486  & 81.786 & 83.329 \\
10.678 & 46.83 & 3.871 &  3.391 & 1.494 & 46.831 & 53.851 \\
10.679 & 13.17 & 5.058 & 3.395 & 1.475  & 73.357 & 76.158 \\ 
14.896 & \phantom{1}9.43  & 4.955 &  3.420 &  1.472  & 69.371 & 73.142\\
14.896 & 50.57 & 4.955  &  3.423 &  1.472  & 69.430 & 73.145 \\
\hline
36.047 & \phantom{1}3.89$^{(a)}$ & 4.836 & 3.432 & 1.475 & 63.895 & 68.508 \\
36.047 & \phantom{1}3.89$^{(b)}$ & 4.613 & 3.429 & 1.482 & 56.128 & 61.914 \\
\end{tabular}
\end{ruledtabular}
\end{table}

\begin{figure}
\includegraphics[width=3.0in]{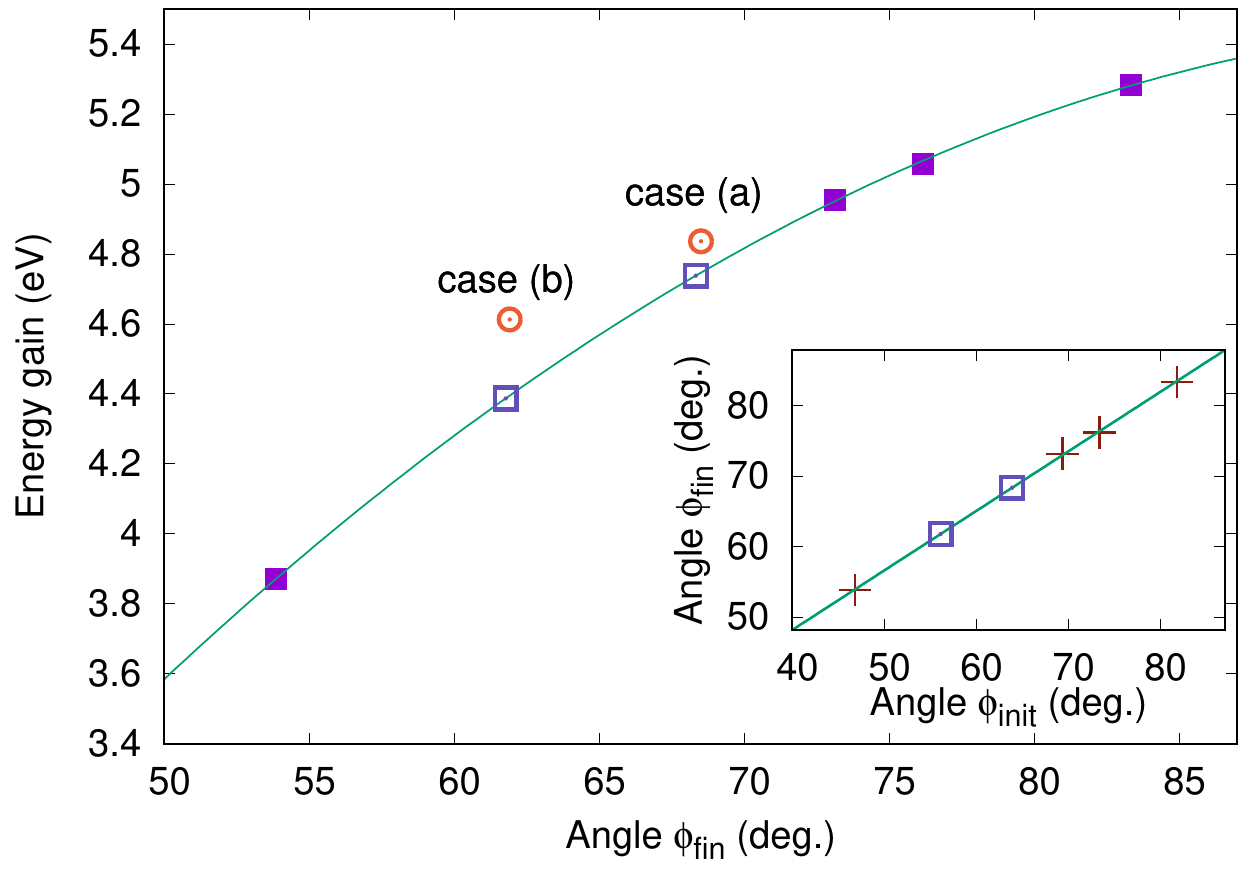}
\caption{\label{fig:en_ang} Energy gain from intercalation in TBG as a function of angle $\phi_{\rm fin}$ (post-relaxation). Filled squares correspond to top 5 rows (including 2 overlapping angles) shown in Table~\ref{table:compcell}. The solid line is a second-order polynomial fit through these 5 cases. The open squares (circles) correspond to the predicted (calculated) energy values for the TBG (3.89\deg) and are labeled according to the last 2 rows in Table~\ref{table:compcell}. A linear fit through the post- and the pre-relaxation angles ($\phi_{\rm fin}$ and $\phi_{\rm init}$) are shown in the inset. The open squares on the straight line show the $\phi_{\rm init}$ values used to predict $\phi_{\rm fin}$, which are then used in the main plot to predict energy (open squares in the main plot) of the TBG (3.89\deg).}
\end{figure}

When the two graphene layers are rotated arbitrarily to give rise to TBG, the size of the supercell is typically large, consisting of a large number of atoms. Compared to 4 atoms in a unit cell of an AA- or AB-stacked bilayer graphene, the smallest supercell in TBG consists of 28 atoms which occurs at angles of twist $\theta\approx 38.21$\deg and $\theta\approx 21.79$\deg \cite{meleprb}. With change in the angle of rotation, the size of the TBG unit cells increases rapidly with a concomitant increase in the number of atoms in the cell. This restricts the scope of using {\it ab initio} calculations to study large TBG unit cells.
However, we show here that the underlying physics behind the unconventional behavior can be understood based on our calculations on smaller TBG unit cells, which can then be generalized to TBG structures with larger unit cells. In the remainder of the paper we refer to specific TBG unit cells as ``TBG ($\theta$)'', where $\theta$ is the angle of twist characterizing the system. 
Note that the size of the TBG unit cells does not vary monotonically with $\theta$, which follows from geometry of a moir\'{e} lattice \cite{melereview,nori}. Therefore, in  Table~\ref{table:compcell}, we list both the lattice parameter of the unit cells as well as the angle of twist to characterize different TBG structures.

We begin with TBG (38.21\deg). We have examined seven unique intercalation positions [shown with orange circles in Fig.~\ref{fig:c_int}(a)], in addition to the ``tetrahedral'' position (marked with green in the same figure), as the initial configurations for the DFT calculations. After ionic relaxation, most pronounced changes in energy and interlayer separation are noted for the case where the C intercalant (henceforth referred to as \Cint\ ) has been placed in, or moved to, the tetrahedral environment. The \Cint\ shows tetrahedral bonding, as it pulls in two carbon atoms from each of the top and bottom layers towards itself [Fig.~\ref{fig:c_int}(b)]. This results in reducing the interlayer spacing locally, and over the entire TBG in this particular case. We are not aware of any previous intercalation cases where the layers are pulled together, even locally. Other configurations of TBG (38.21\deg) show no such dramatic effects, and the second highest energy gain is smaller by more than 1~eV compared to the tetrahedral configuration. These cases are not considered further in this study. On the other hand, TBG (21.79\deg) does not have similar tetrahedral environment to begin with [Fig.~\ref{fig:c_int}(c)], and placing \Cint\ in that structure fail to recreate similar bonding behavior. Placing the \Cint\ in the nearest approximation of tetrahedral environment in TBG (21.79\deg) leads to large-scale ionic rearrangement during the ionic relaxation step, rendering {\it ab-initio} prediction infeasible.

Next, we explore TBG with larger unit cells than the previous two cases. In Table~\ref{table:compcell} we describe geometrical and computed quantities, and the stability energy associated with tetrahedral C intercalation for all the studied TBG structures. The energy gain from intercalation is defined as $\Delta E = -\{E_{n+1}({\rm TBG}+ $\Cint$)\ - E_n({\rm TBG}) - E^{\rm C, ref}\}$. Here $E_{n+1}({\rm TBG}+ $\Cint) and $E_n({\rm TBG})$ refer to the total energy of the TBG cases with and without the intercalant, respectively, and $E^{\rm C, ref}$ corresponds to the energy of a single carbon atom in a reference structure and can be set to 0 without affecting the relative energy-gain values. The biggest $\Delta E$ is observed for TBG (38.21\deg) described earlier, which also shows a staggering reduction of over 20\% in the average interlayer separation to 3.157~\AA~from the no-intercalation separation of 3.41~\AA.
On the other hand, the smallest energy gain is found for TBG (46.83\deg) at 3.871~eV. The other cases show intermediate energy gains, and their average interlayer separation are within $\sim$0.05~\AA\ of the no-intercalation cases. The bond lengths between \Cint~and layer atoms in all these cases are within 0.025~\AA\ of each other, despite the much larger variation seen in the average interlayer separation. In other words, in all the cases that we studied, the interlayer separation is 2.5-2.6~\AA\ locally around the \Cint.
In order to specify the local tetrahedral geometry we introduce the angle $\phi$, defined as the angle between the upper and the lower tetrahedral planes, formed by the \Cint\ (common to both planes) and the two layer atoms residing in the upper and the lower graphene layers, respectively [see Fig.~\ref{fig:c_int}(a)]. In Table~\ref{table:compcell} we note both $\phi_{\rm init}$ and $\phi_{\rm fin}$, corresponding to the angle between the tetrahedral planes before and after ionic relaxation. We expect that a bonding environment with $\phi$ closer to 90\deg\ is energetically favored, and all the cases considered in Table~\ref{table:compcell} support this picture.

In Figure~\ref{fig:en_ang} we show the correlation between $\Delta E$ and $\phi_{\rm fin}$, with the solid line showing a second-order polynomial fit. We also find that the values of $\phi_{\rm fin}$ are linearly related to $\phi_{\rm init}$ to a high degree of accuracy (Fig.~\ref{fig:en_ang} inset). This suggests that the $\phi_{\rm init}$ values (pre-ionic-relaxation values from TBG geometry) could be used to predict $\phi_{\rm fin}$, which can then be used to predict $\Delta E$. To test the accuracy of this prediction, we estimate the post-relaxation energy gain purely based on the geometry, for two tetrahedral environments occurring in TBG (3.89\deg), having 868 C atoms in total (Fig.~\ref{fig:m3-89}). The corresponding computed values of $\Delta E$ are presented in Table~\ref{table:compcell}: The energy gain estimates are found to be within 0.1~eV for case (a) and 0.2~eV for case (b). 

Typical TBG samples in experiments have unit cells of the order of a few nm, containing a large number of atoms \cite{van2}. In such large lattices, the atomic registry changes much more slowly over large length scales. This is already apparent in Fig.~\ref{fig:m3-89}, a unit cell consisting of 868 atoms. If one follows the direction in which the atomic registry changes smoothly from locally AA- to AB- to BA- and back to AA-like regions, it can be seen that in between AB- and BA-like regions, the atomic arrangements are very close to being tetrahedral. We expect the appearance of such tetrahedral bonding environment in these transition regions to be a persistent feature as one goes to larger unit cells in the small-twist-angle regime, and we visually verified this for a number of large TBG unit cells. In such structures, \Cint\ will prefer to place itself in these transition regions, owing to larger $\Delta E$.  

It is instructive to consider what happens when one intercalates conventional AA- or AB-stacked bilayer instead of TBG with carbon. From geometry, AA- or AB-bilayer does not offer any local environment that would allow tetrahedral bonding with C atoms intercalated between the layers; therefore, we do not expect the same behavior as in TBG. Indeed, calculations initiated with a \Cint\ placed halfway between the layers of AA- or AB-bilayer, end up with the intercalant being adsorbed to one of the layers. Our first-principles calculations on 3$\times$3 supercells of AB (AA)-structures indicate that such adsorption configurations are associated with an energy gain of 2.842 (3.304)~eV. For comparison, we also performed a calculation where we started with the adsorbed configuration for the AA-structure (minus the intercalant), and placed the intercalant halfway between the layers. When constrained in this configuration, we found an energy gain of 1.069~eV. This smaller energy gain explains the strong preference towards adsorption over bonding with both layers for these cases. Thus, local environments provided by the moir\'{e} superlattice in TBG are necessary to observe the unusual behavior reported here.

\begin{figure}
\includegraphics[width=3.0in]{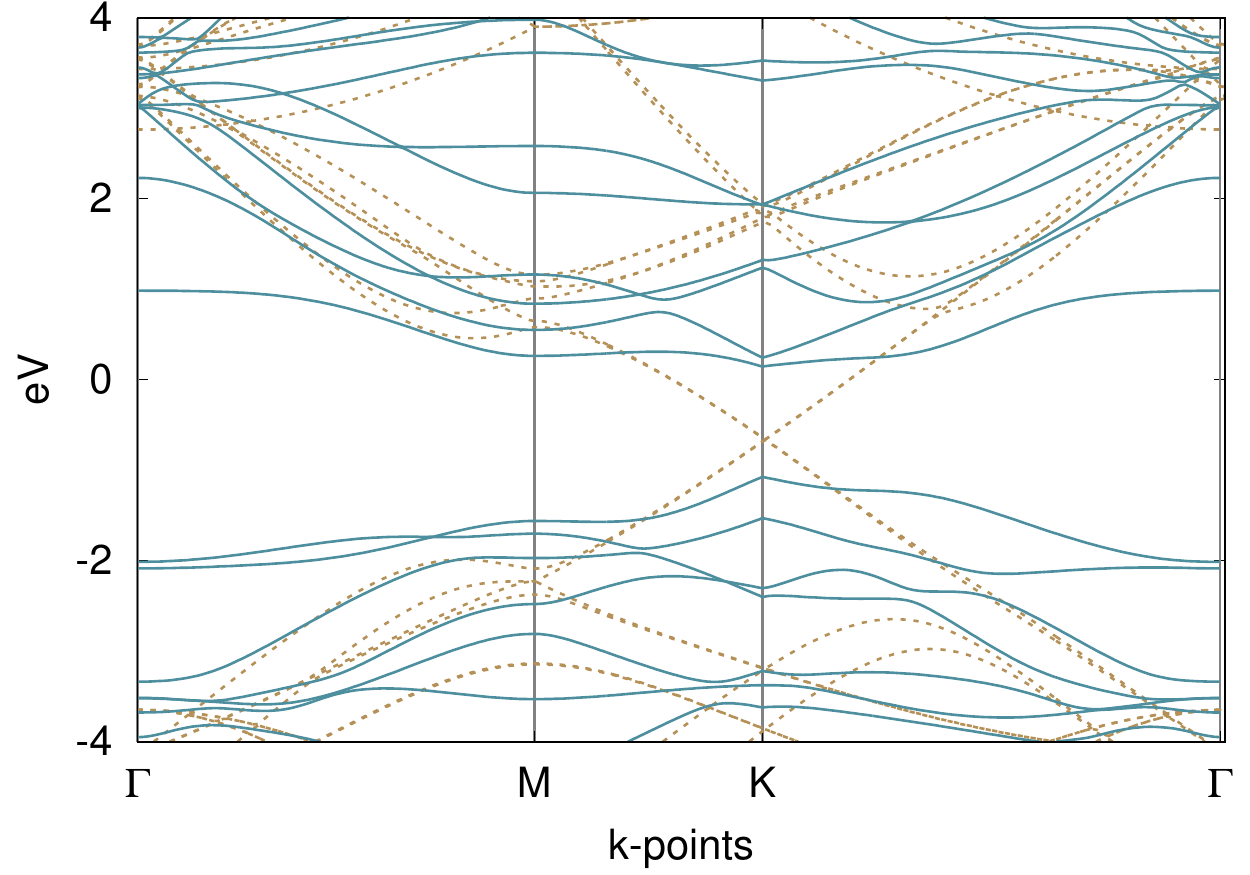}
\caption{\label{fig:m38-bs} (Color online) Band structure of TBG (38.21\deg) before (dashed lines) and after intercalation (solid lines). Intercalation introduces a large gap at the Dirac point.}
\end{figure}

The structural modification described above also affects the electronic spectrum. As an example, we show in Fig.~\ref{fig:m38-bs} the band structure of TBG (38.21\deg) both before and after carbon intercalation. As can be seen, a substantial gap opens up at the Dirac point, in addition to introducing significant distortion of the bands of pristine TBG (38.21\deg). The opening of the gap is expected and has been reported before in cases where TBG was intercalated with other intercalants\cite{bandgap}.

To summarize, we find that carbon as an intercalant is capable of bonding tetrahedrally with host carbon atoms in TBG  when the latter provide such an environment (quantified via the introduction of $\phi$ in this work) within the lattice. The bonding with the intercalant results in the two layers being pulled closer to each other locally in the intercalation regions.
The nontrivial behavior requires TBG as a host: AA- and AB-cases do not show such behavior since they do not have a suitable tetrahedral environment.
Additionally, it is important that C is used as an intercalant to observed the nontrivial behavior. Although we do not know of any other intercalant bonding simultaneously with both layers of TBG, intercalants such as N and O are known to form chemical bonds \cite{gong_first-principles_2010} with single graphene layers or one of the layers in TBG. We cannot rule out the possibility of such elements bonding with both layers if local environments permit. Using our calculations on TBG lattices, we also show how the energy gain $\Delta E$ from intercalation could be approximated from the geometry of pristine (i.e., no intercalation) TBG structures, including cases of large unit cells in the small-twist-angle regime.

We expect the chemical bonding of C$_{\rm int} $ with both graphene layers to result in clamping the layers together, leading to a loss of the lubrication property that ordinarily exists in graphene and other 2D systems. On the other hand, interlayer bonding would provide structural stability in all three directions. The tetrahedral bonding environment of the C$_{\rm int}$ raises the possibility of other potential intercalants, made of organic motifs such as $sp^2$-hybridized C$_2$ backbone or suitably substituted organic compounds, to be used similarly for intercalation. By inserting different functional organic compounds, it may be possible to  controllably vary the interlayer separation in TBG, with potential application as sensors and for battery electrodes. Further, our study points to the possibility of such non-trivial behavior in other van der Waals heterostructures as well.

Downloadable data files related to this work will be made available after peer review.

\begin{acknowledgements}
AR acknowledges support from NSF Grant No.~DRL-1237992 and from Extreme Science and Engineering Discovery Environment (XSEDE) computation research grant (Grant No.~TG-DMR160019). XSEDE is supported by NSF Grant No.~ACI-1053575. HKP acknowledges support by the French program ANR DIRACFORMAG (ANR-14-CE32-0003) and LabEx PALM Investissement d'Avenir (ANR-10-LABX-0039-PALM).
\end{acknowledgements}

\end{document}